\documentclass[11pt]{article}

\usepackage{amssymb,amsthm,amsmath}
\usepackage{graphicx}
\usepackage[total={6.5in,8.75in}, top=1.2in, left=0.9in, includefoot]{geometry}

\def\be{\begin{equation}}
\newcommand{\bel}[1]{\begin{equation}\label{eq:#1}}
\def\ee{\end{equation}}

\def\bea{\begin{eqnarray}}
\def\eea{\end{eqnarray}}
\def\ba{\begin{align*}}
\def\ea{\end{align*}}

\def\pp{p_{\phi}}

\def\al{\alpha}

\def\pr{\partial}
\def\haf{\frac {1}{2}}
\newcommand{\R}{{\mathbb{ R}}}
\def\HH{H\'enon-Heiles~}

\newtheorem{thm}{Theorem}
\newtheorem{lem}[thm]{Lemma}
\newtheorem{cor}[thm]{Corollary}

\theoremstyle{definition}

\newcommand{\Eq}[1]{(\ref{eq:#1})}
\newcommand{\Th}[1]{Th.~\ref{thm:#1}}
\newcommand{\Lem}[1]{Lem.~\ref{lem:#1}}
\newcommand{\Cor}[1]{Cor.~\ref{cor:#1}}

\newcommand{\Sec}[1]{\S \ref{sec:#1}}
\newcommand{\Fig}[1]{Fig.~\ref{fig:#1}}

\newcommand{\InsertFig}[4]
{\begin{figure}[ht]
       \centerline{
         \includegraphics[width=#4]{#1.pdf} 
       }
       \caption{{\footnotesize  #2}
       \label{fig:#3}}
\end{figure}}

\newcommand{\InsertFigTwo}[5] {
\begin{figure}[ht]
       \centerline{
         \includegraphics[width=#5]{#1.pdf} 
         \hskip 0.5in
         \includegraphics[width=#5]{#2.pdf}
       }
       \caption{{\footnotesize  #3}
       \label{fig:#4}}
\end{figure}}

\newcommand{\InsertFigThree}[6] {
\begin{figure}[ht]
       \centerline
       {
           \includegraphics[height=#6]{#1.pdf} 
           \hskip 0.25in
           \includegraphics[height=#6]{#2.pdf}
           \hskip 0.25in
        	   \includegraphics[height=#6]{#3.pdf}
       }
       \caption{{\footnotesize  #4}
       \label{fig:#5}}
\end{figure}}

\title{Straight Line Orbits in Hamiltonian Flows}
\author{  
       J.E. Howard$^{1}$ and J.~D.~Meiss$^{2}$
{     
}    
\smallskip\\
\smallskip\\
	$^{1}$Laboratory for Atmospheric and Space Physics\\
		and Center for Integrated Plasma Studies\\
		University of Colorado\\ 
		Boulder, CO 80309, USA\\
		{\tt Jhoward@Colorado.edu}
\smallskip\\
$^{2}$Department of Applied Mathematics\\
      University of Colorado \\
      Boulder, CO 80309-0526, USA\\
      {\tt James.Meiss@colorado.edu} 
}

\begin{document}
\maketitle

\begin{abstract}
We investigate periodic straight-line orbits (SLO) in Hamiltonian force fields using both direct and inverse methods.  A general theorem is proven for natural Hamiltonians quadratic in the momenta in arbitrary dimension and specialized to 
two and three dimension.  Next we specialize to homogeneous potentials and their superpositions, including the familiar H\'enon-Heiles problem. 
It is shown that SLO's can exist for arbitrary finite superpositions of $N$-forms. 
The results are applied to a family of generalized H\'enon-Heiles potentials having discrete rotational symmetry. SLO's are also found for 
superpositions of these potentials. 
\end{abstract}


\section{Introduction}

The connection between the geometry of trajectories and the force fields that generate them has long been of interest to physicists and mathematicians alike \cite{Szebehely67,Bozis95,vanderMerwe91}. In the copious literature on the subject one may distinguish between \emph{direct} and \emph{inverse} approaches. The  direct problem asks: what are closed form solutions for particular orbits in a given potential? By contrast, the inverse problem poses the question: which force field will produce a set of orbits with a given shape?
For example, suppose that the ``natural" Hamiltonian, $H: \R^{2n} \to \R$,
\bel{natural}
	H(q,p) = \tfrac12 |p|^2 + V(q) \;,
\ee
has an orbit with energy $E$ that lies on a given surface in the configuration space,
\[
	F(q,c) = f(q) - c = 0 \;.
\]
where $f: \R^n \to \R$. As was first shown by Szebehely \cite{Szebehely74} for the planar case, and generalized by Puel \cite{Puel84} to $n$-dimensions, the potential must satisfy the equation
\bel{Puel}
	\frac{\nabla f}{|\nabla f|} \cdot \nabla V = 2(E-V) \nabla \cdot 
			\left( \frac{\nabla f}{|\nabla f|} \right)
\ee
on $F = 0$. If the orbit is restricted to a curve, given by the intersection of $n-1$ surfaces $F_i = f_i(q) - c_i$, then the potential must satisfy \Eq{Puel} for each $i=1, \ldots n-1$.  The potentials obtained in this way have an $n-1$ parameter family of orbits lying on the intersection of the sets $F_i = 0$; these equations have been studied in a number of papers and many three dimensional examples have been obtained \cite{Puel88, Bozis04, Bozis05}.

In contrast to Szebehely's problem, we consider the problem of finding a potential with a \emph{single} orbit of a given shape. In lieu of specifying this curve as the intersection of surfaces, we find it more convenient to represent it parametrically through $Q: \R \to \R^n$, as 
\bel{parametric}
	q(t) = Q(x(t)) \;,
\ee
where $x(t) \in \R$ now represents the temporal dynamics. Note that $x-Q(x) = 0$ may be regarded as a {\it configurational invariant} \cite{Hall83} valid for one particular value of the energy $E$. 

Substituting this into the equations of motion $\ddot q = - \nabla V$ gives
        $Q' \ddot x + Q'' \dot x^2 = - \nabla V$.
Since $|p|^2 = |Q'|^2 \dot x^2 = 2(E-V)$, we can eliminate the first derivative to obtain
\[
        Q' \ddot x=  - \nabla V - \frac{2(E-V)}{|Q'|^2} Q''.
\]
The implication is that the vector on the right must be in the $Q'$ direction, in other words that the projection onto the plane orthogonal to the tangent vector $Q'$ must be zero. The projection matrix orthogonal to the vector $Q'$ is
\[
        P = \left( I - \frac{1}{|Q'|^2}Q' Q'^T \right)
\]
(note that $P^2 = P$ and $P Q' =0 $).
We then obtain
\bel{curve}
        P \left (\nabla V + \frac{2(E-V)}{|Q'|^2} Q'' \right) = 0
\ee
along $q = Q(x)$. If $V$ satisfies \Eq{curve} then the dynamics reduces to the scalar system
\bel{scalar}
        \ddot x = - \frac{ Q'}{|Q'|^2}\cdot \nabla V(Q) - 2 \frac{Q' \cdot Q''}{|Q'|^4} (E-V(Q)) \;.
\ee
Thus the inverse problem reduces to finding a potential that satisfies \Eq{curve}. In general this seems to be a hard problem that we will leave to a later paper.

The simplest geometry for an orbit is a straight line in the configuration space:
\bel{SLOform}
	q(t) = Q(x(t)) = s x(t) + q_0 \;,
\ee
for a constant ``slope vector" $s$, i.e., a straight line orbit (SLO). In this case, \Eq{curve} simplifies considerably since $Q'' = 0$, and the requirement on the potential is simply that its gradient must be parallel to $Q' = s$, or specifically
\bel{SLO}
	 \nabla V|_{s x + q_0} = \lambda(x) s \;,
\ee
for some scalar function $\lambda$. In this case the dynamical equation \Eq{scalar} reduces to
\bel{SLOdyn}
	\ddot x = -\lambda(x) \;.
\ee
Since \Eq{SLO} is independent of the energy, the straight line orbits that we find automatically come in one-parameter families, parametrized by $E$.


Our investigations possess some of the aspects of both direct and inverse methods. Indeed, for a given potential $V$, we can solve \Eq{SLO} to determine the allowed values of $s$, if any. We will call the set of admissible slopes, the \emph{slope spectrum} of $V$. Alternatively, we can fix $s$ and solve the eigenvalue-like problem \Eq{SLO} for the potential; the general solution of \Eq{SLO} will be obtained in \Sec{spectrum}.
Subsequently, we will specialize to the two and three degree-of-freedom cases, giving a number of examples.  For two degrees of freedom, the general solution to \Eq{SLO} involves two arbitrary functions, and for three, it involves four functions. Another class of examples, superpositions of homogeneous potentials, is treated in \Sec{homogeneous}. An example of this case is the famous \HH system which has three families of SLOs. By expressing this case in polar coordinates we also obtain SLOs for a family of hyper-\HH potentials.

In all cases we are motivated by physical applications and decline interest in discovering exotic potentials which will never be found in a physical problem. 

The ideas here should be contrasted with the notions of central configuration and choreography in celestial mechanics. A central configuration is a solution in which $\ddot q = -\lambda(q) q$, for some scalar function $\lambda: \R^n \to \R$ \cite{Moeckel90}. When the masses are equal, the potential must satisfy $\nabla V = \lambda(q) q$, instead of \Eq{curve} or \Eq{SLO}. A choreography is a solution of an $N$-body system in which \emph{each} body has identical configuration space, and each follows the same curve, but with a phase shift \cite{Chenciner02}. In the standard gravitational problem, the configuration space is $\R^{3N}$ for $N$ bodies in $\R^3$ and each body has a configuration orbit that lies on a curve $q_i(t) = C(x(t) + \phi_i)$ for a curve $C: \R \to \R^3$.

\section{Straight Line Orbits}\label{sec:spectrum}
As in the introduction, we consider a Hamiltonian system on $\R^{2n}$ with coordinates $(q,p)$.  A straight line orbit has the form \Eq{SLOform} with intercept $q_0$, slope vector $s \in \R^n$ and scalar dynamical function $x(t)$. For any particular SLO we can, without loss of generality, choose coordinates so that $q_0 = 0$; consequently, we will look only for orbits that go through the origin. Moreover, since the equation is homogeneous in $s$, we can choose the slope vector so that $|s| = 1$.

In particular consider the \emph{natural} Hamiltonian system \Eq{natural} with potential $V \in C^2(\R^n,\R)$. If $H$ admits an SLO with slope $s$, then 
\begin{align*}
	p &= s \dot x \\
	s \ddot x &= - \nabla V(q)|_{q = sx} \;.
\end{align*}
Thus an SLO exists only when $\nabla V(sx)$ is parallel to $s$ for all $x \in \R$. We call the set of admissible slopes the \emph{slope spectrum}, $Sl(H)$ of the Hamiltonian $H$. The slope spectrum for \Eq{natural} is thus determined by a nonlinear eigenvector-like equation
\bel{slopeSpectrum}
	Sl(H) = \{ s \in \R^n: \nabla V(sx) = \lambda(x) s\;, \lambda: \R \to \R\}.
\ee

The general form of a potential admitting an orbit with a given slope can be easily determined:

\begin{thm}\label{thm:SLOGen}
The Hamiltonian system \Eq{natural} has a family of straight line orbits $q(t) = sx(t)$, $|s| = 1$, only if the potential has the form
\bel{potentialForm}
	V(q) = U(s^T q) + \tfrac12 q^T W(q) q \;,
\ee
where $W: \R^n \to \R^{n \times n}$ is any symmetric matrix function that has a zero eigenvector $s$, $W(q) s = 0$. In this case $x: \R \to \R$ is any solution of the one-dimensional ODE
\bel{oneDegree}
	\ddot x = -U'(x) \;.
\ee
\end{thm}

\begin{proof}
It is not hard to see that \Eq{potentialForm} satisfies \Eq{slopeSpectrum}. To show this is the general form, we  choose a new basis aligned with $s$. Let $R = (s,M)$ be an $n \times n$ orthogonal matrix, so that the columns of the $n \times (n-1)$ matrix $M$ are orthonormal and orthogonal to $s$: $M^T s = 0$. The slope requirement \Eq{slopeSpectrum} then becomes the system of $n-1$ equations: $M^T \nabla V(sx) = 0$.

Defining new coordinates by $q = R \xi = sx + My$, where $\xi = (x,y) $ and $y \in \R^{n-1}$, then the potential in the new coordinates is $\tilde U(x,y) = V(sx + My)$. Noting that $\nabla_q = R^{-T} \nabla_\xi = R \nabla_\xi$, the slope equation becomes
\[
	\left. M^T R \nabla_{(x,y)} \tilde U(x,y)\right|_{y = 0} = 0 \;.
\]
But since $R$ is orthogonal, 
\[
	I = R^T R = \begin{pmatrix} s^T R \\ M^TR \end{pmatrix}\;,
\]
so $M^T R = (0,I)$ (where this $I$ has size $n-1$) and the slope equation reduces simply to
\bel{criticalpoint}
	\left.\partial_{y} \tilde U(x,y)\right|_{y = 0} = 0 \;.
\ee
This is just the requirement that $U$ has a zero derivative with respect to the $n-1$ variables $y$ when they vanish. It has general solution
\[
	\tilde U(x,y) = U(x) +  \tfrac12 y^T \tilde W(x,y) y
\]
where $\tilde W$ is a smooth, symmetric $(n-1) \times (n-1)$ matrix function. In terms of the original coordinates, note that $y = M^T q$, and $x = s^T q$ so that
\[
	V(q) = \tilde U(R^T q) = U(s^T q) + \tfrac12 q^T M \tilde W M^T q \;.
\]
Note that the matrix $W = M \tilde W M^T$ is symmetric and that its rank is no more than $n-1$; it has a zero eigenvector $s$. Thus we have the general solution \Eq{potentialForm}.  The equation of motion on the SLO immediately reduces to the one-degree-of-freedom system \Eq{oneDegree}.
\end{proof}

If $V$ has a local minimum at the origin, then \Eq{oneDegree} will have some bounded solutions. However, this does not imply that the resulting orbits are stable when thought of as orbits of the full system.

We turn next to some examples for two and three degrees of freedom.

\subsection{Two-degree-of-freedom natural flows}\label{sec:2DOF}

Here we give an explicit form for \Eq{potentialForm} for the case of two degrees of freedom assuming a straight line orbit of the form $q(t) = (x(t), \alpha x(t))$. In this case, the requirement on the potential reduces to the single equation 
\bel{spectrum2D}
	Sl(H) = \left\{\alpha: \left(\alpha \frac{\partial}{\partial x}- 
	      \frac{\partial}{\partial y} \right) V(x,\alpha x) = 0 \right\} \;,
\ee
which could easily be solved directly. However, it is also a simple application of \Th{SLOGen}.

\begin {cor}\label{thm:PDE}
The natural Hamiltonian \Eq{natural} with $(q,p) = (x,y,p_x,p_y)$ has a straight line orbit of the form $y(t) = \alpha x(t)$ only if
\bel{potential2DOF}
	V(x,y) = F \left({x+\alpha y}\right) + (\alpha x-y)^2 G(x,y) \;,
\ee
where $G$ is continuous at $y=\alpha x$. In this case $x$ obeys the ODE $\ddot x = -F'((1+\alpha^2) x)$.
\end{cor}

\begin{proof} For this case $s \propto (1, \alpha)$ and the coordinate transformation is
\[
	R = (s,M) 
	  = \frac{1}{\sqrt{1+\alpha^2}} \begin{pmatrix} 1 &-\alpha \\ \alpha & 1 \end{pmatrix}
\]
and the $2 \times 2$ matrix $W$ in \Eq{potentialForm} has $s$ as a zero eigenvector when it is proportional to $MM^T$, giving
\[
	W(q) = 2F(x,y) \begin{pmatrix} \alpha^2 & -\alpha \\ -\alpha & 1 \end{pmatrix}
\]
Thus \Eq{potentialForm} becomes
\[
	V(q) = U(s^Tq) + (\alpha x - y)^2 F(x,y).
\]
Scaling the argument of $U$ gives \Eq{potential2DOF}
\end{proof}

For example
\[
	V(x,y) = \cos(x+y) + (x-y)^2 \sin x \cos y
\]
has orbits $y(t) = x(t)$ with $x$ obeying the pendulum equation
\[
	\ddot x = \sin 2 x.
\]
It is also easy to construct potentials which have multiple straight line orbits. For
example:
\[
	V(x,y) = (x-y)^2 F(x,y) \;,
\]
has $x=y$ as an orbit. We now may replace $F$ by a function that has other straight lines; for example,
\[
	V(x,y) = (x-y)^2 x^2 (x+y)^2
\]
has $y=x$, $x=0$ and $y=-x$ as orbits.

\subsection{Three-dimensional natural flows}\label{sec:3DOF}

Consider the three degree of freedom case of \Eq{natural} and a straight line orbit of the form $L = \{(x(t),\alpha x(t), \beta z(t))\}$. The slope spectrum is then given by 
\bel{spectrum3D}
	Sl(H) = \left\{(\alpha,\beta): 	\left.\frac{\partial}{\partial x} V \right|_{L} 
	   = \left.\frac{1}{\alpha}\frac{\partial}{\partial y} V \right|_{L} 
	   = \left.\frac{1}{\beta} \frac{\partial}{\partial z} V \right|_{L} \right\}\;.
\end{equation}
This could be solved directly, but it is also easy to directly apply \Th{SLOGen}.

\begin{cor}\label{cor:3D}
The natural Hamiltonian \Eq{natural} with $(q,p) = (x,y,z, p_x,p_y,p_z)$ has a family of straight line orbits $q(t) = (x(t),\alpha x(t), \beta x(t))$ only if functions $A(x,y,z)$, $B(x,y,z)$, $C(x,y,z)$ and $U(x)$ exist such that
\begin{align*}
	V(x,y,z) = U(x+\alpha y+\beta z) + (\alpha x -y)^2 A + (\beta x -z)^2 B + (\beta y - \alpha z)^2 C \;,
\end{align*}
where $A$ is continuous at $y=\alpha x$, $B$ at $z =\beta x$ and $C$ at $\beta y = \alpha z$.

\end{cor}

\begin{proof}
Here we set $s \propto (1,\alpha,\beta)$. Requiring that the matrix $W$ in \Eq{potentialForm} has $s$ as a zero eigenvector yields the form
\[
	W = 2 \begin{pmatrix} 
	   \alpha^2 A+\beta^2 B   &-\alpha A        &-\beta B\\
	   -\alpha A              &A+\beta^2 C      &-\alpha \beta C\\
	   -\beta B               &-\alpha \beta C  &B+\alpha^2C
		\end{pmatrix} \;.
\]
This results in the quadratic form
\[
	q^T W q = (\alpha x -y)^2 A + (\beta x -z)^2 B + (\beta y - \alpha z)^2 C
\]
which immediately gives the result.
We will give several examples in Sec. 4.
\end{proof}

\section{Homogeneous Potentials}\label{sec:homogeneous}

Homogeneous potentials are often encountered practice, and we are therefore motivated to develop a method specifically tailored for this class. In particular suppose that $H$ has the form
\bel{polynomialH}
	H = \haf |p|^2 + \sum_{k=2}^N U^{(k)}(q) \;,
\ee
where each term $U^{(k)}(\alpha q) = \alpha^k U^{(k)}(q)$ is homogeneous with degree $k$.
We shall concentrate on the two and three degree of freedom cases.
Examples include the \HH system \cite{Henon64} and its generalizations \cite{Hall83}, where the potentials are polynomial.

Note that both \Eq{spectrum2D} and \Eq{spectrum3D} reduce to individual equations for each homogeneous term in $U$:

\begin{lem}\label{lem:Summation}
The slope spectrum for \Eq{polynomialH} is the intersection of the slope spectra for the homogeneous potential system $H^{(k)} =\tfrac12 |p|^2 + U^{(k)}(q)$. For any $s \in Sl(H)$, with $|s| = 1$,  the orbit $q(t) = s x(t)$ satisfies 
\bel{oneDegreeII}
	\ddot x = - \sum_{k=2}^N  kU^{(k)}(s) x^{k-1}
\ee
\end{lem}

\begin{proof}
The requirement of \Eq{slopeSpectrum} becomes
\[
	\sum_{k=2}^N (\nabla U^{(k)}(s) - \Lambda_{k-1} s)x^{k-1} = 0 \;,
\]
where $\lambda(x) = \sum_{k=1}^{N-1} \Lambda_k x^k$ is also a polynomial in $x$.  Each term in the sum is a homogeneous polynomial in the scalar function $x(t)$ of degree $k-1$, and, unless $x(t)$ is constant, these terms must vanish individually since different powers of a non-constant function are linearly independent. This gives the individual ``eigenvalue" problems
\bel{HomoEV}
	\nabla U^{(k)}(s) = \Lambda_{k-1} s.
\ee
Thus	 
\[
	Sl(H) = \bigcap_{k=2}^N Sl(H^{(k)}) \;.
\]

As in the general case, for each $s \in Sl(H)$, $x$ must solve \Eq{SLOdyn}. By \Eq{HomoEV},
$ \Lambda_{k-1} = s \cdot \nabla U^{(k)}(s)$, which by homogeneity becomes $\Lambda_{k-1} = k U^{(k)}(s)$. Thus \Eq{SLOdyn} reduces to \Eq{oneDegreeII}.
\end{proof}

Specializing now to the case of polynomials, we consider first the 2D case where
\bel{UN}
	U^{(N)}(x,y) = \sum_{n=0}^k a_n x^n y^{N-n} \;. 
\ee
If we look for an orbit with slope $s = (1,\alpha)$, the slope spectrum requirement \Eq{spectrum2D} reduces to
\bel{QN}
	Q^{N}(\al) \equiv \sum_{n=1}^N [n a_n \al -(N-n+1)a_{n-1}]\al^{N-n} = 0 \;.
\ee
This equation can be thought of in two ways. For a given $\alpha$, it can be viewed as a single linear restriction on the coefficients. Alternatively, for a given set of coefficients, \Eq{QN} becomes a single polynomial equation in $\alpha$ whose real zeros determine the slope spectrum.

In particular the quadratic case 
\[
	Q^{(2)}(\alpha) = a_1 \alpha^2 +2(a_2-a_0)\alpha - a_1
\]
always has two real zeros since its discriminant
\[
	\Delta = (a_2-a_0)^2 + a_1^2
\]
is nonnegative. There is one special case, in which any $\alpha$ is in the slope spectrum: $Q^{(2)}$ is identically zero when $a_1 = 0$ and $a_2 = a_0$, which corresponds to the harmonic oscillator case
\[
	U_{SHO} = \frac12 (x^2 + y^2) \;.
\]

The cubic case reduces to
\[
	Q^{(3)}(\alpha) = a_1\alpha^3+ (2a_2-3a_0) \alpha^2+ (3a_3-2a_1) \alpha -a_2.
\]
This, of course, has at least one real zero, so the slope spectrum is always nonempty.
%

It is also easy to find examples of homogeneous, but non-polynomial potentials with SLO's. For example, for the degree-two potential
\[
	U^{(2)}(x,y) = \frac{9x^4 +8x^3y+5y^4}{x^2+y^2}
\]
we obtain
\[
	Q^{(2)}(\alpha) = 4\frac{(2-\alpha)(5\alpha-1)(1+\alpha)}{1+\alpha^2}
\]
which has three real zeros giving the slope spectrum
\[
	Sl(H) = \{ 2 ,-1, \tfrac15\}
\]

Now consider the three degree of freedom case, with
\be
	U^{(N)}(x,y,z) = \sum_{m+n+q=N} a_{nmq} x^n y^m z^q.
\ee
The requirement \Eq{spectrum3D} reduces to the two equations
\bel{QN3D}
\begin{split}
	Q^{(N)}_1(\al,\beta) &= \sum a_{nmq} (m\al^{m-1} - n\al^{m+1}) \beta^q = 0 \;,\\
	Q^{(N)}_2(\al,\beta) &= \sum a_{nmq} (q\beta^{q-1} - n\beta^{q+1}) \al^m = 0 \;.
\end{split}
\ee
Again these equations can be viewed in two ways. For a given pair $(\alpha,\beta)$ they are a set of simultaneous linear equations for the coefficients $a_{mnq}$. Alternatively, for given coefficients, the two polynomials many have a set of simultaneous solutions that give the slope spectrum. These solutions may be found by taking the resultant of $Q_1$ and $Q_2$.

 
\section {Examples}\label{sec:examples}

\subsection {The \HH System}
The well-studied \HH system is a natural Hamiltonian with potential \cite{Henon64}
\bel{HH}
	V(x,y) = \tfrac12(x^2+y^2) + x^2y -\tfrac13 y^3 \;.
\ee
The standard $(y,p_y)$ section with $x = 0$ and $p_x >0$ for total energy $E=H=\tfrac18$ is shown in \Fig{HHsection}. 

\InsertFig{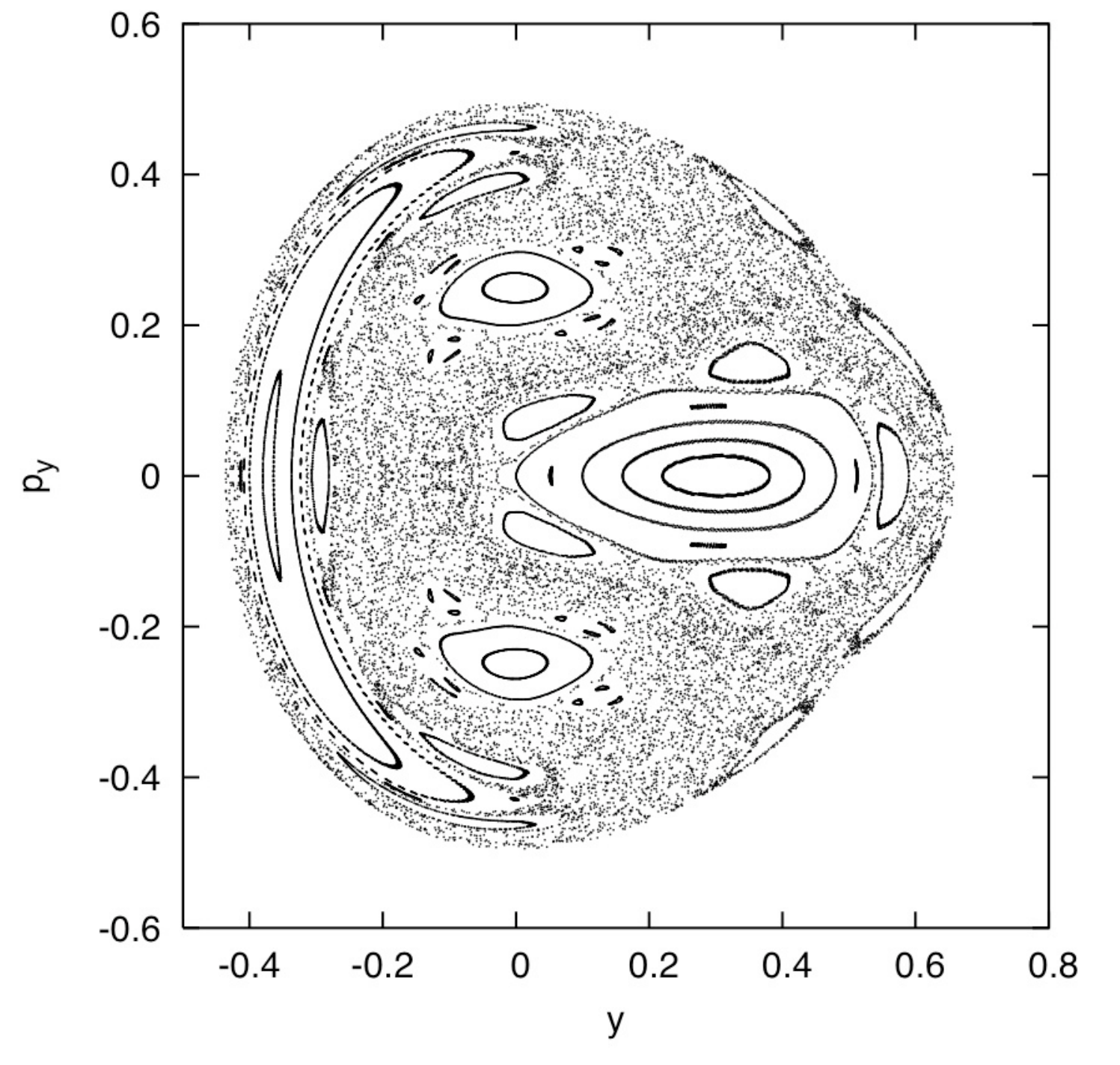}{Poincar\'e section for \HH potential, with $E=1/8$}{HHsection}{3in}

To find straight line orbits for the potential \Eq{HH}, we can apply \Lem{Summation}. Since the quadratic part is symmetric, it has SLOs for any $s$. The slope spectrum \Eq{slopeSpectrum} for the cubic part gives the eigenvector equation
\[
	\begin{pmatrix} 2s_1 s_2 \\ s_1^2-s_2^2\end{pmatrix} = \Lambda
	\begin{pmatrix} s_1 \\ s_2\end{pmatrix}
\]
which reduces to the equations $s_1 = 0$ and $s_2 = -\Lambda$, or $s_1^2 = 3 s_2^2$ and  $2s_2 = \Lambda$. Thus the slope spectrum for the \HH Hamiltonian is
\[
	Sl(H) = \{(0,1), (\sqrt3,1),(-\sqrt3,1) \} \;.
\]
These three straight line orbits are shown projected onto configuration space in \Fig{HHorbits}(a). Once we know the slopes, we can show that \Eq{HH} is of the form \Eq{potential2DOF} by first solving for $F$ using $F\left((1+\alpha^2)x\right) = V(x,\alpha x)$, and then solving for $G$. For example for $\alpha = \frac{1}{\sqrt3}$ we find
\begin{align*}
	F(x) &= \frac{\sqrt3}{8}x^2 (x+\sqrt3) \;, \\
 	G(x,y) &= \frac38 ( 1-\sqrt3x - y) \;.
\end{align*}

The SLO orbits with  $\alpha = \pm 3^{-1/2}$ appear as fixed points on the $y-$axis in the section of \Fig{HHsection}. These orbits cross the section $x=0$ at $y=0$ and since $p_y = \alpha p_x$, this implies that the momentum on the section is
\bel{py0}
	p_{y0} = \pm \sqrt{\frac{2\alpha^2 E}{1+\alpha^2}}\;.
\ee
For $E = \tfrac18$, the SLOs are elliptic and form the centers of the two stable islands at $p_y = \pm \tfrac14$ in \Fig{HHsection}. These fixed points are stable up to $E \approx 0.14$, at which point a pitchfork bifurcation occurs generating a pair of stable periodic orbits that are no longer SLOs; one of these orbits is shown in \Fig{HHorbits}(b). The vertical SLO, with $x\equiv 0$ lies in the section and corresponds to its boundary, namely the contour of
\[
	E = H(0,y) = \tfrac12(p_y^2 + y^2) -\tfrac13 y^3 \;.
\]
Similar results have been obtained previously by Antonov and Timoshkova \cite{Antonov93} and van der Merwe \cite{vanderMerwe91}.

\InsertFigTwo{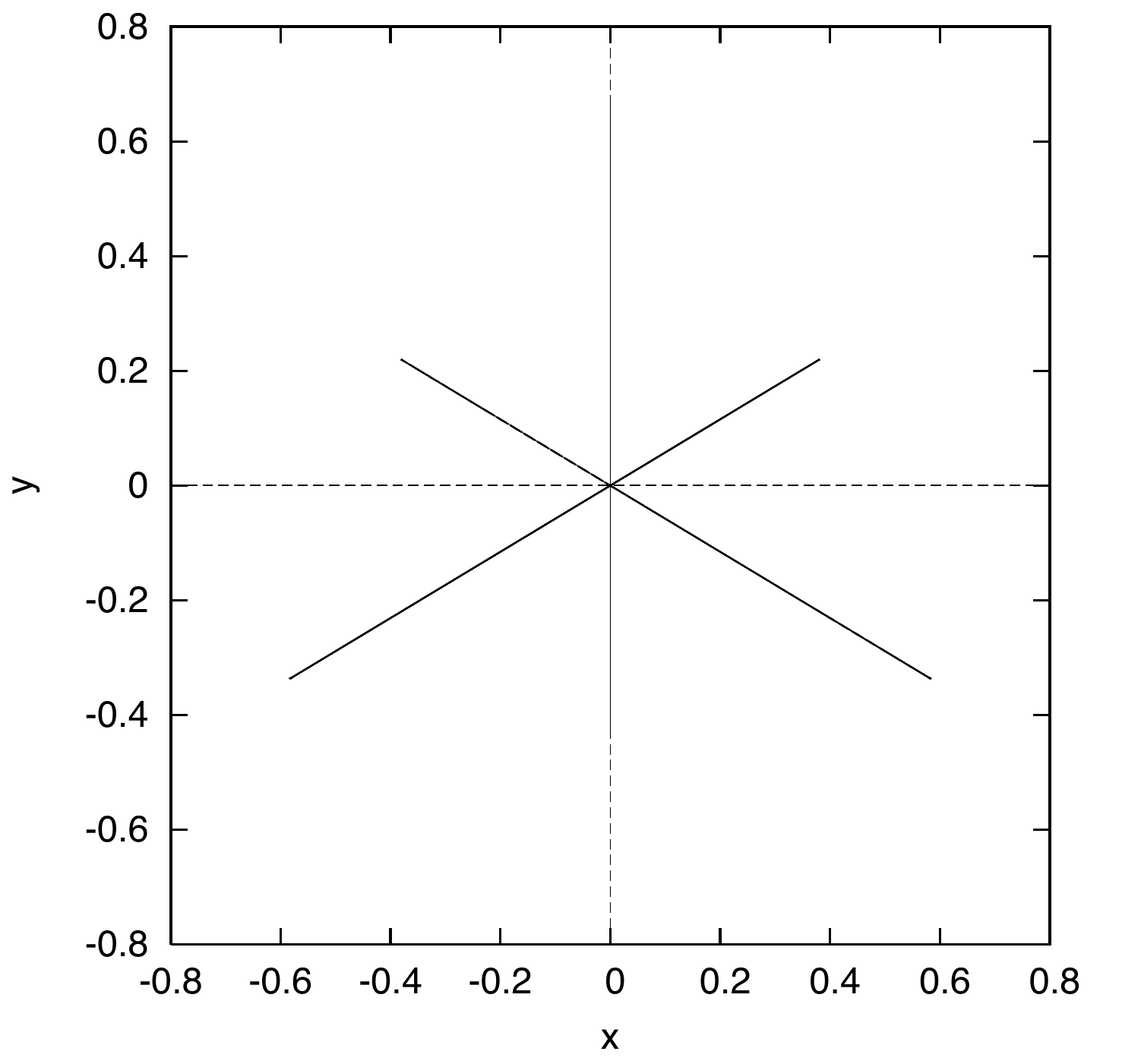}{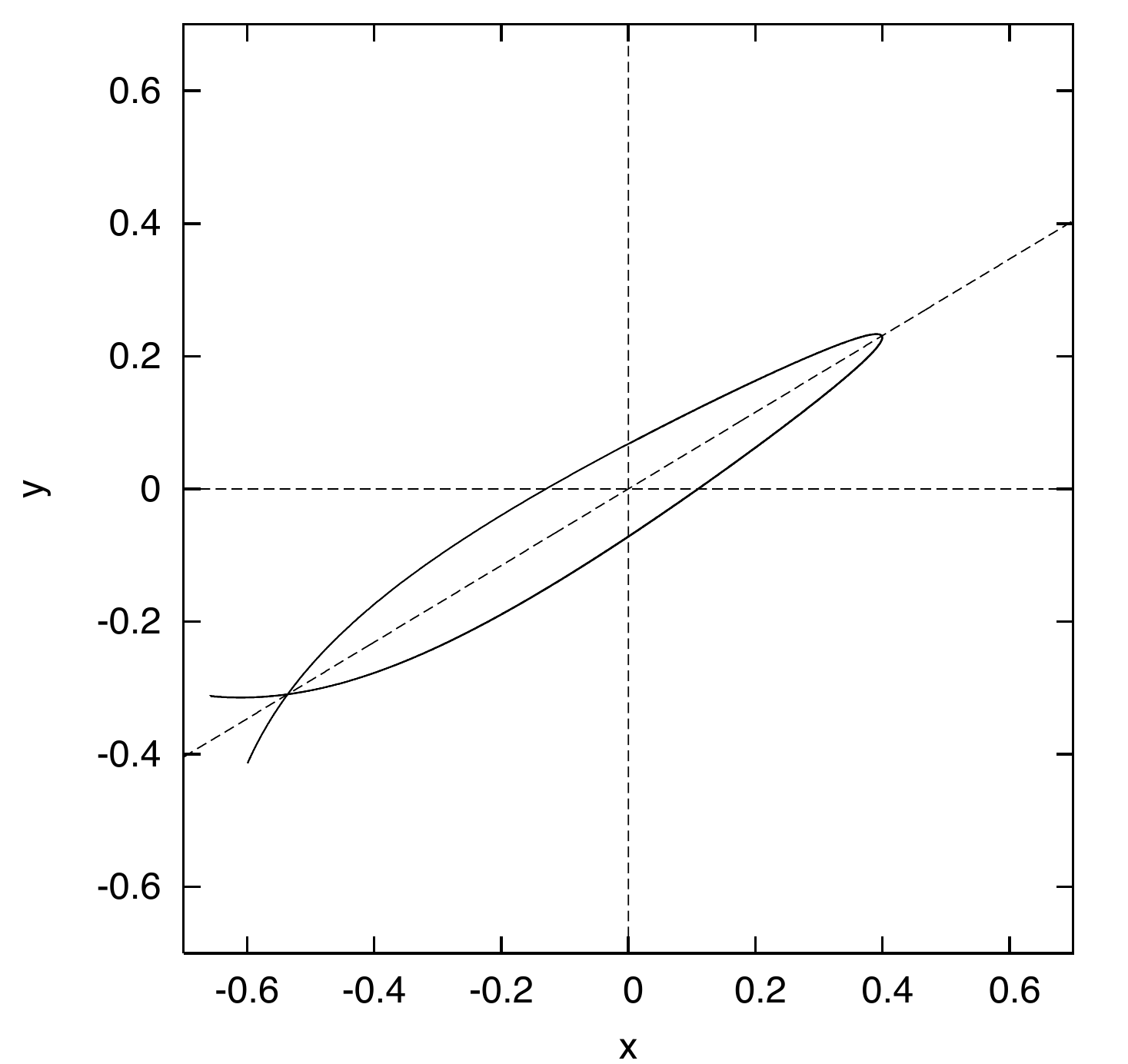}{(a) The three SLOs for the \HH system for $E = 1/8$, (b) stable bifurcated orbit at $E = 0.14$.}{HHorbits}{3in}

\subsection{A Quartic Potential}
As is well known, the potential for the \HH system can be written in polar coordinates as $U^{(3)} = \frac{r^3}{3} \sin 3\phi$.
This suggests the quartic analogue 
\bel{Quartic}
	U^{(4)} = \frac{r^4}{4} \sin 4\phi =   x^3 y -xy^3.
\ee
For this potential \Eq{QN} becomes the biquadratic, 
\[
	Q^{(4)}(\alpha) = \al^4 -6 \al^2 + 1 = 0
\]
with roots $\al^2 = 3 \pm 2\sqrt{2}$  which implies that $\al = \pm \tan{\pi/8}$ and $\pm \tan{3\pi/8}$. The resulting four SLOs are shown in \Fig{QuarticOrbits}(a) for $E = \tfrac18$.

\InsertFigThree{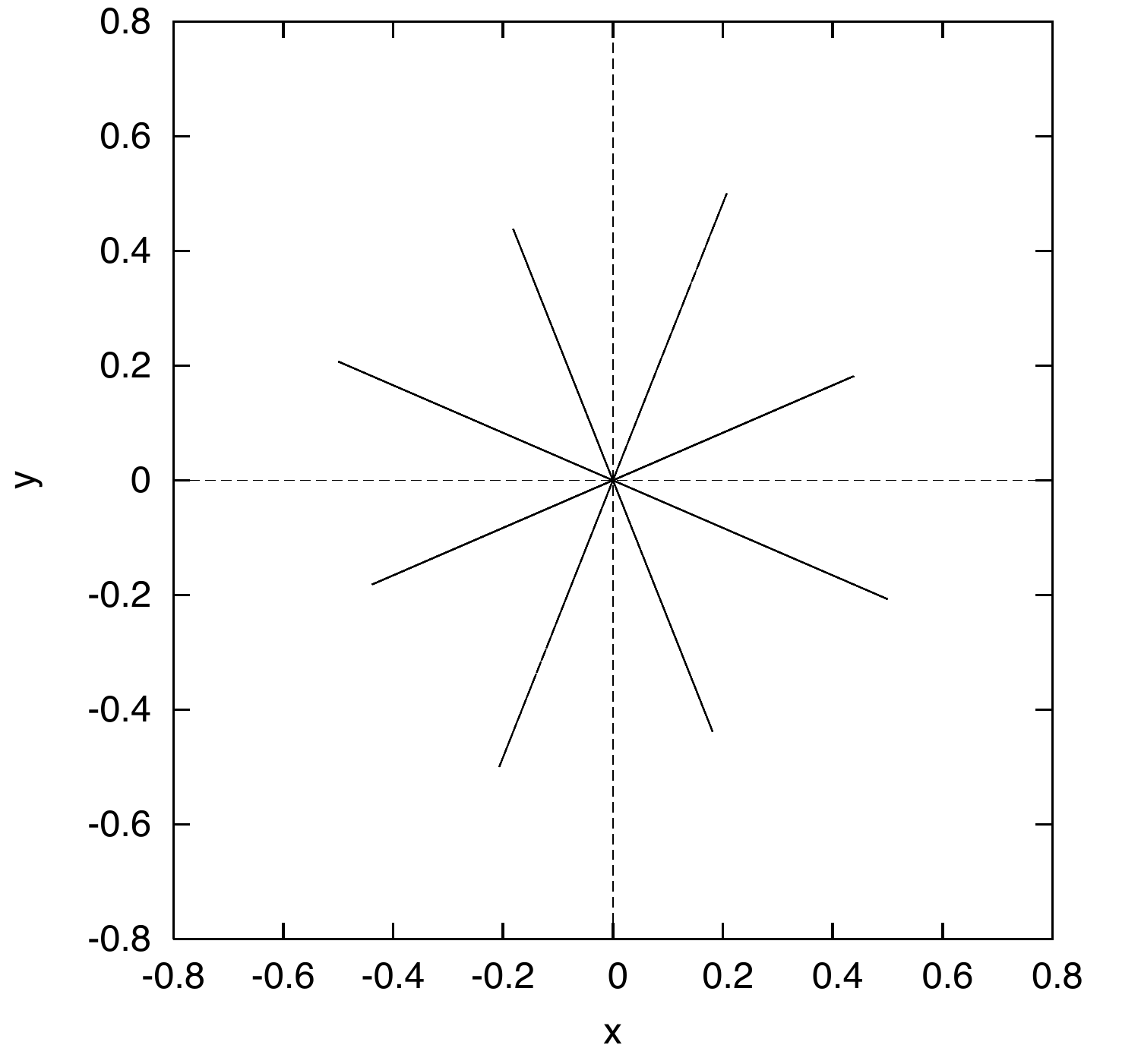}{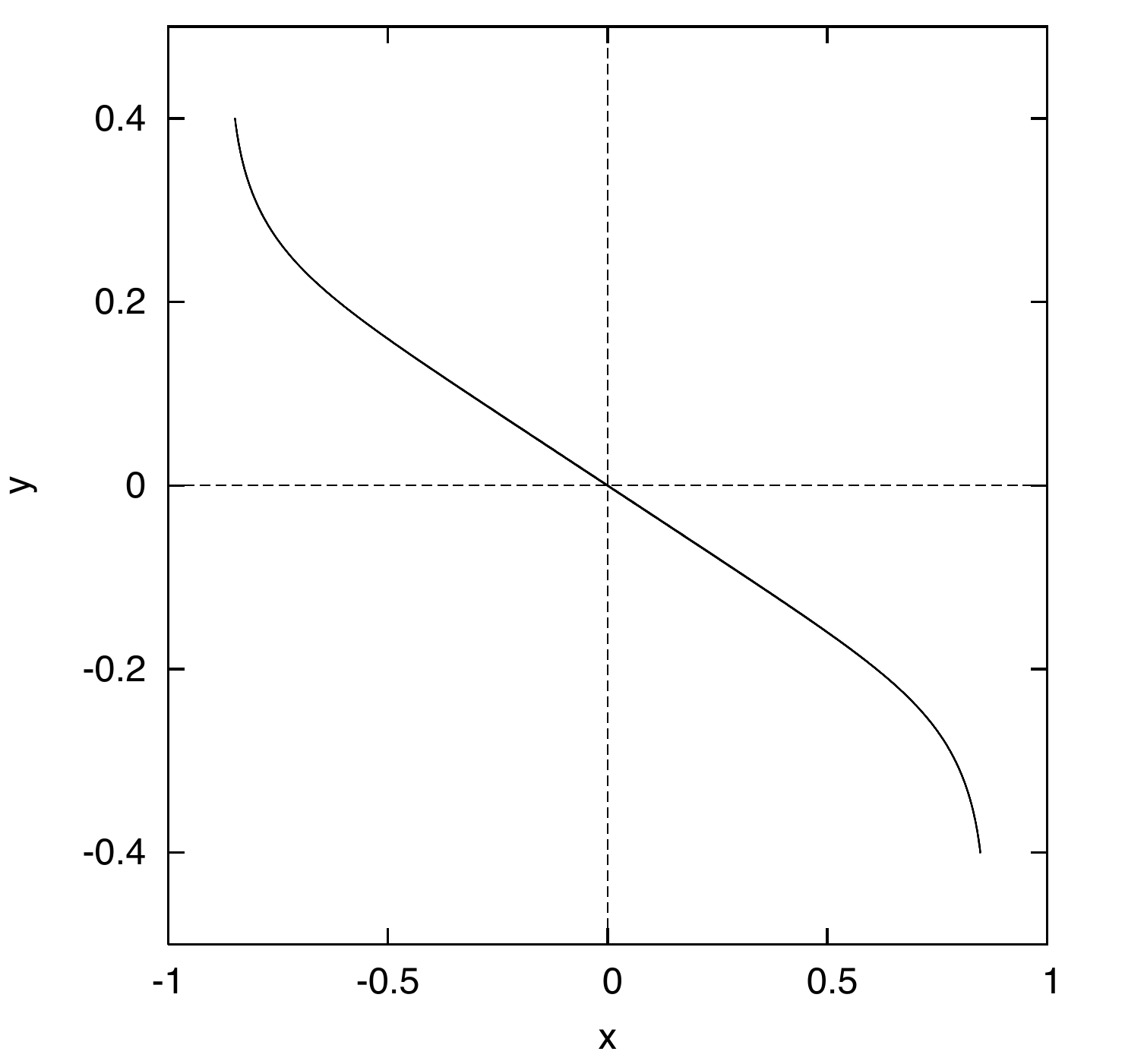}{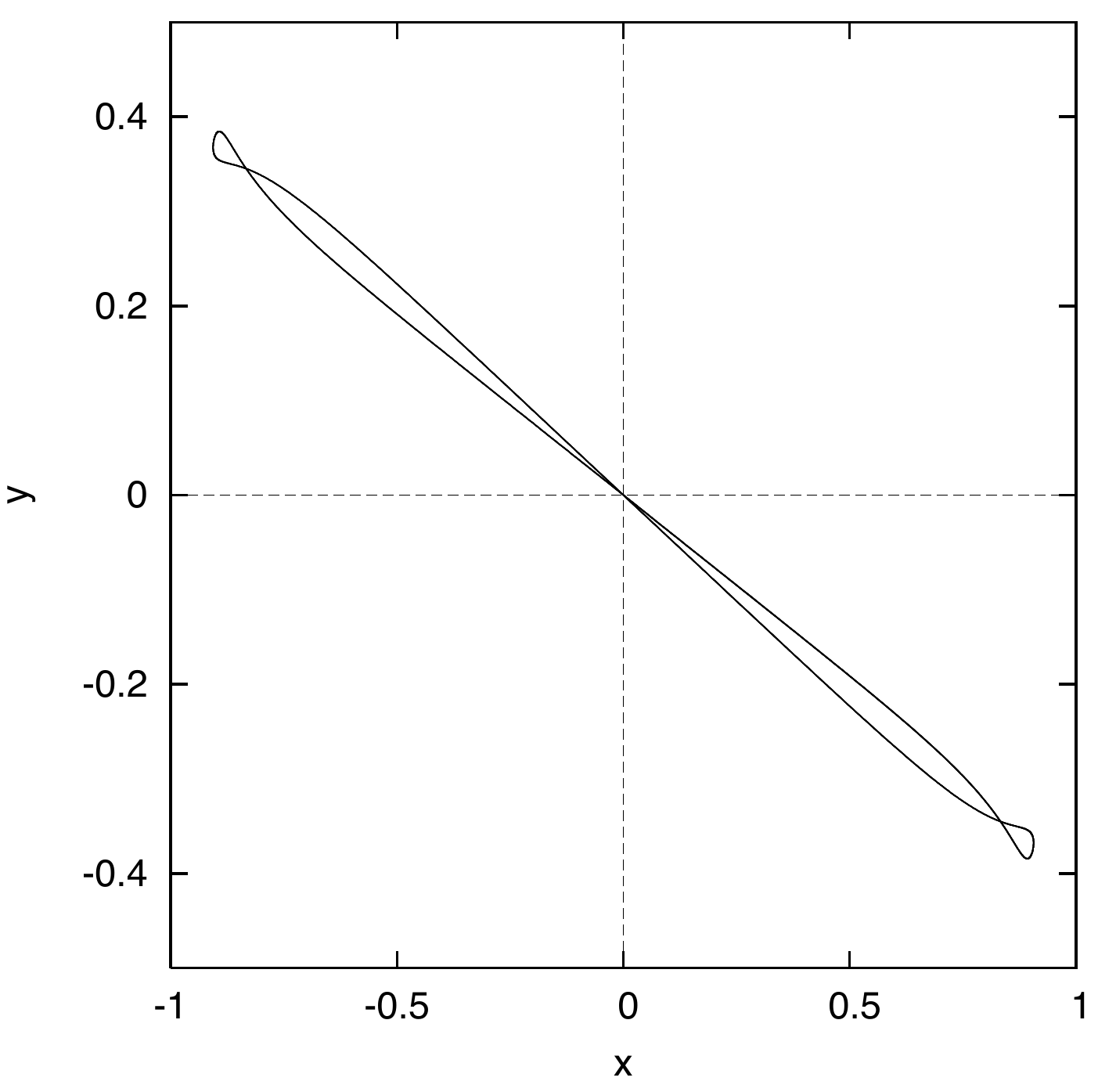}
{(a) The four straight-line periodic orbits for the potential $\frac12 r^2 + U^{(4)}$ with the quartic \Eq{Quartic}, with $E=\tfrac18$, for $\al \approx \pm 0.191341$ and $\pm 0.46194$. (b) period-one orbit for $E=\tfrac14$ at $p_{y0} = -0.213$ (c) period-one orbit at $p_{y0} = -0.2518$.}{QuarticOrbits}{2in}

The section $x=0$, $p_x >0$ is shown in \Fig{QuarticSection}; for $E = \tfrac18$ the four SLOs correspond to elliptic fixed points on the $y$-axis at the momenta determined by \Eq{py0}, namely, $p_{y0} \approx \pm 0.191$ and $\pm 0.462$. As the energy increases, bifurcations occur, resulting in a changing number of fixed points orbits on $y$-axis. These orbits are ephemeral, coming and going with changing $E$. For example at $E = \tfrac14$, the section, shown in \Fig{QuarticSection}(b), there are at least six elliptic fixed points on the $y$-axis.
In \Fig{QuarticOrbits}(b)-(c) two of the additional period-one orbits, located at $p_{y0} \approx -0.213$ and $-0.2518$, are shown. Although these are clearly not SLOs, the first one is remarkably linear near the origin and self-retracing. \Fig{QuarticBifs} depicts a bifurcation sequence that generates additional period-one orbits. For $E=0.21$ there are two SLOs in the lower half-plane; for $E = 0.22$, one SLO has destabilized by a subcritical pitchfork bifurcation, spawning two stable non-SLOs. At $E=0.23$ this SLO has restabilized via a supercritical pitchfork bifurcation. There are now a total of 4 period-one orbits in the lower half plane, of which two are SLOs. Of course, the number of SLO's is constant.

\InsertFigTwo{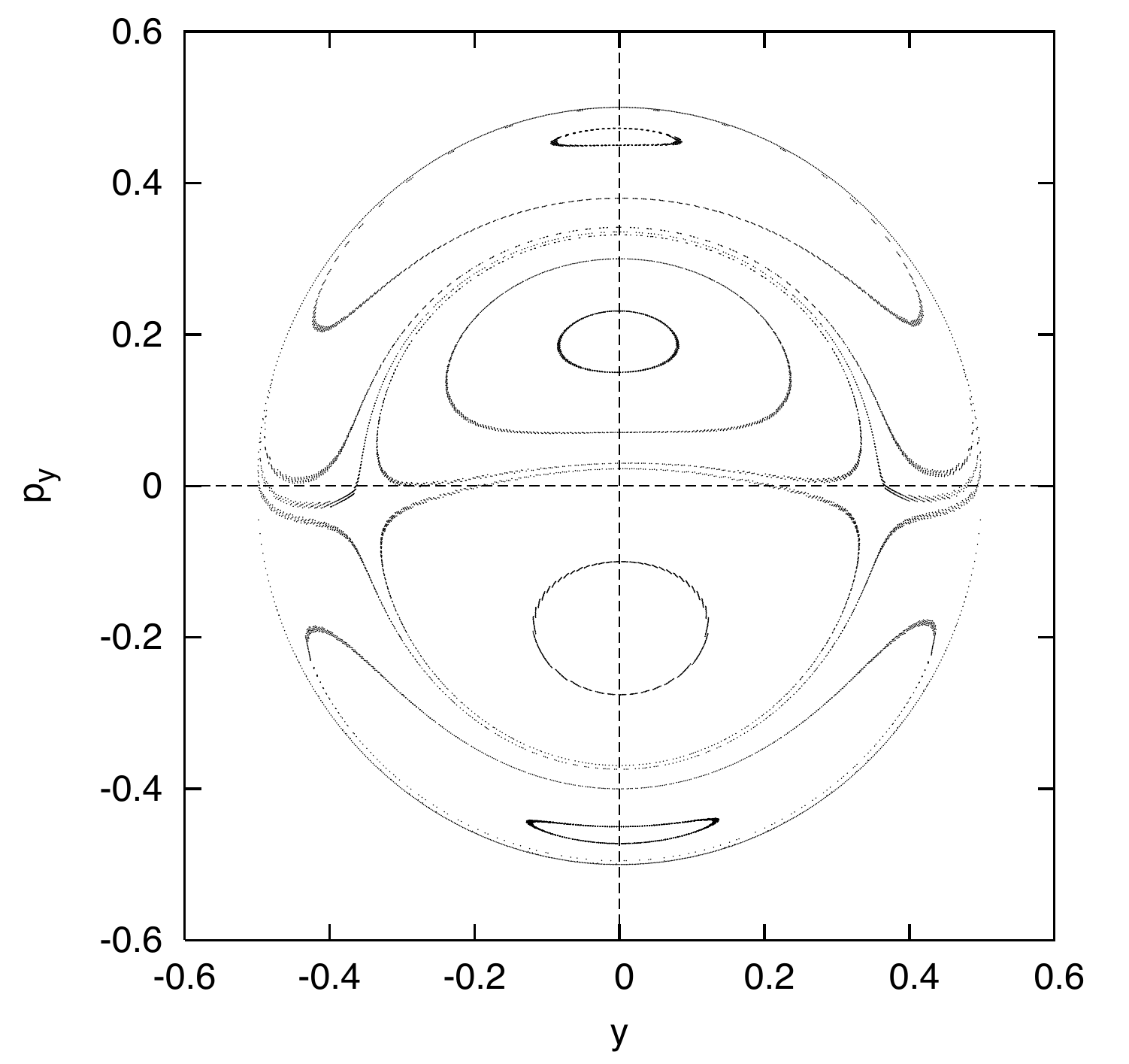}{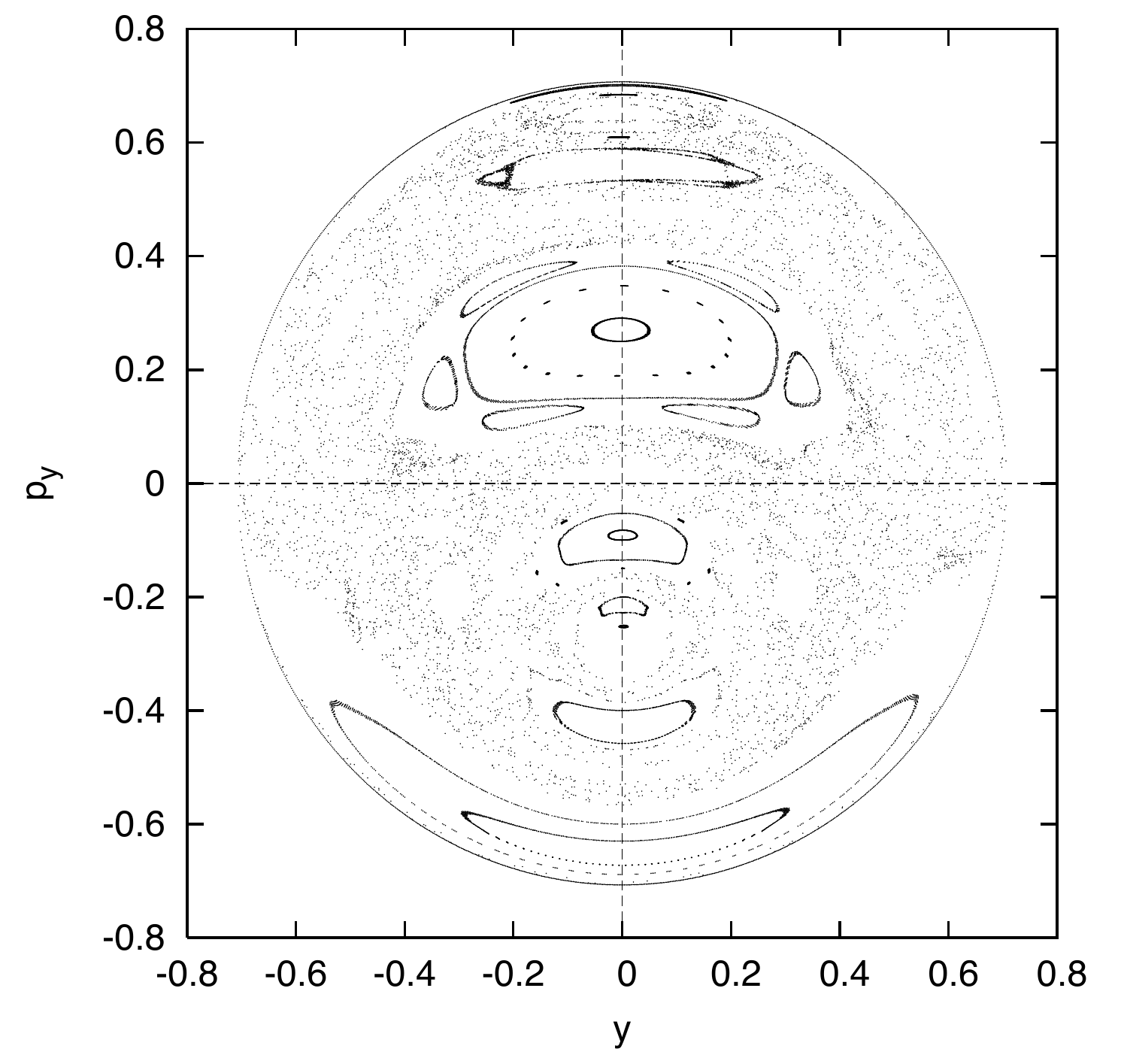}
{Poincar\'e section for quartic potential \Eq{Quartic}, with (a) $E=1/8$, (b) $E = 1/4$.}{QuarticSection}{3in}

\InsertFigThree{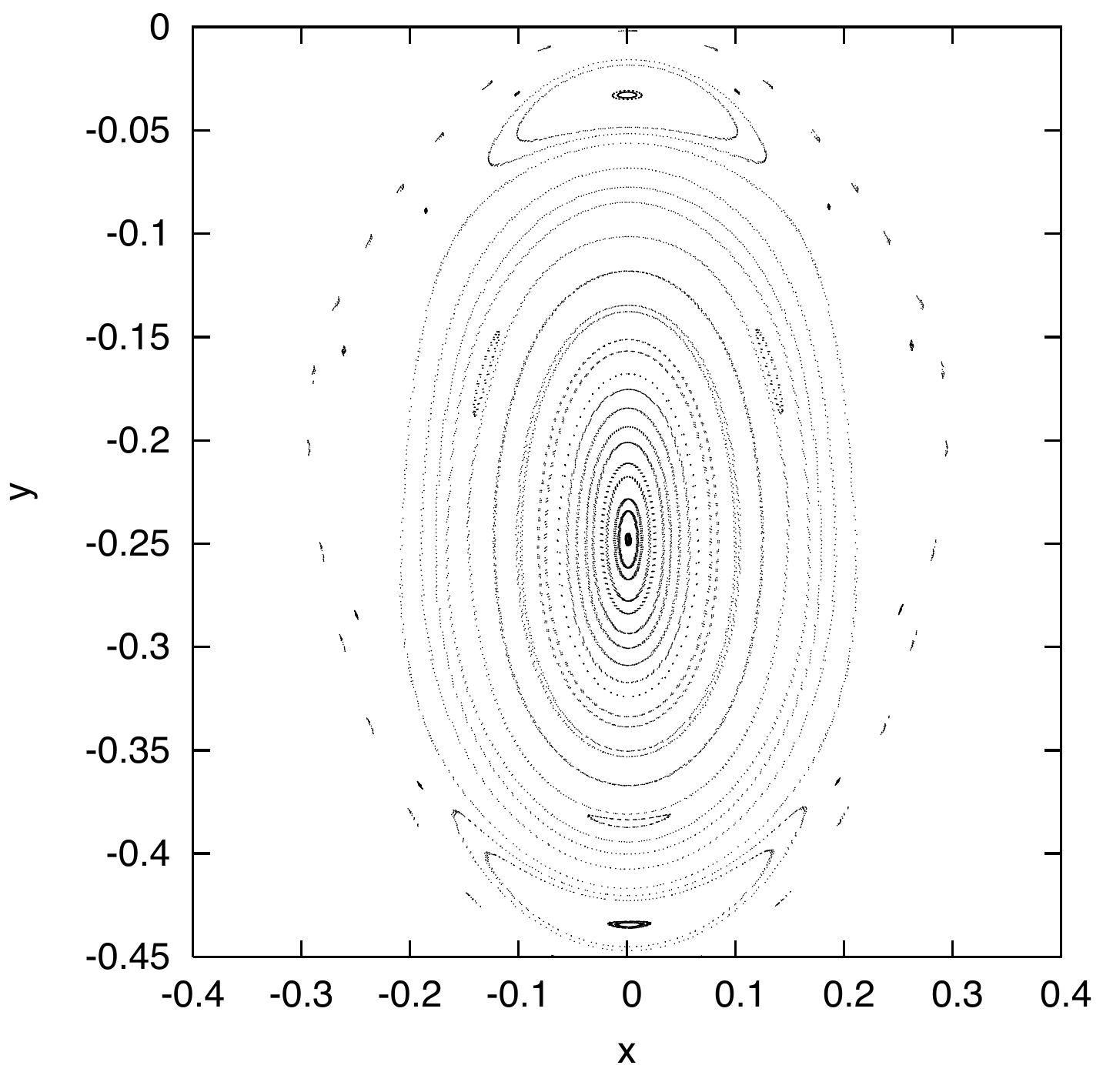}{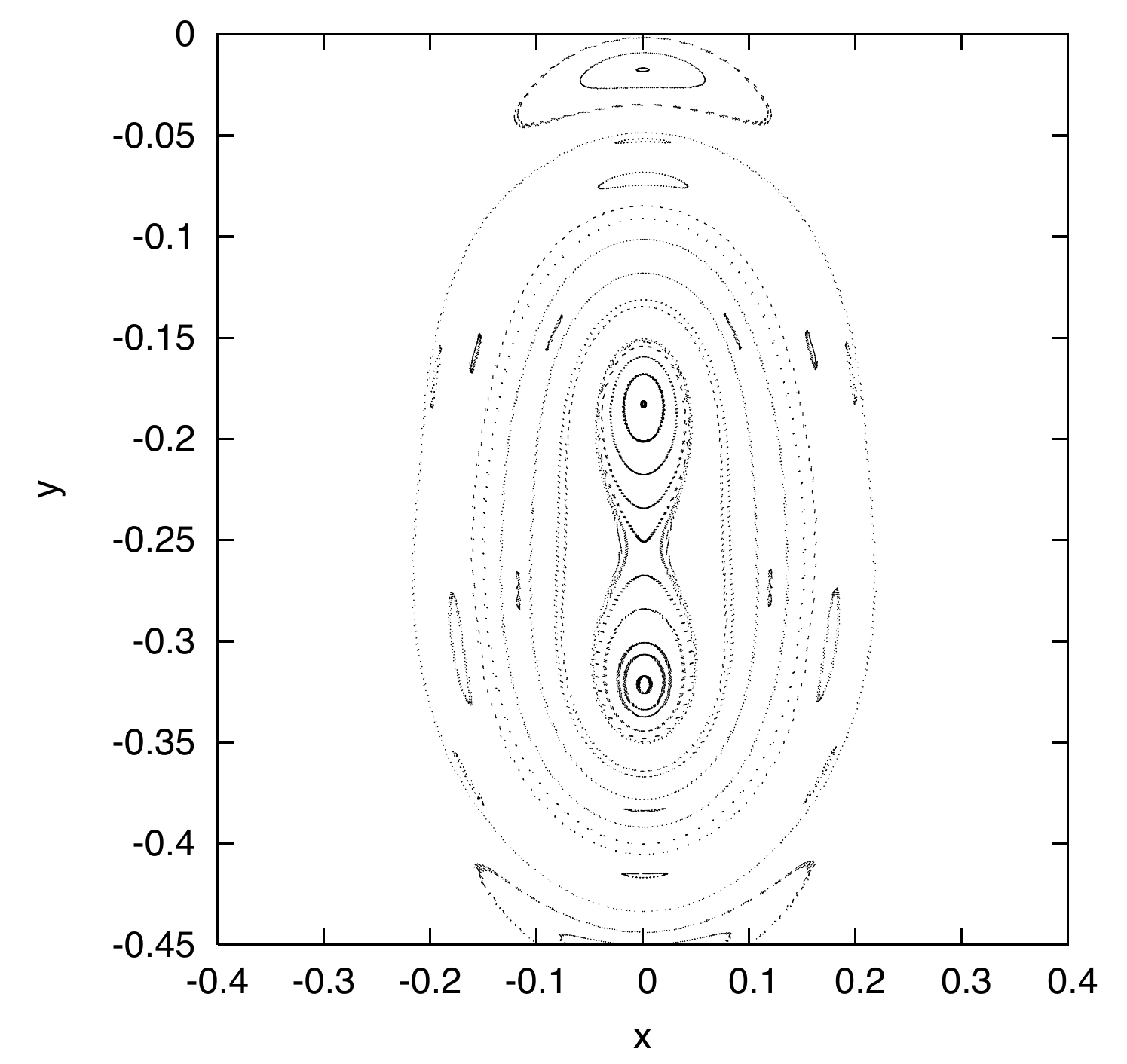}{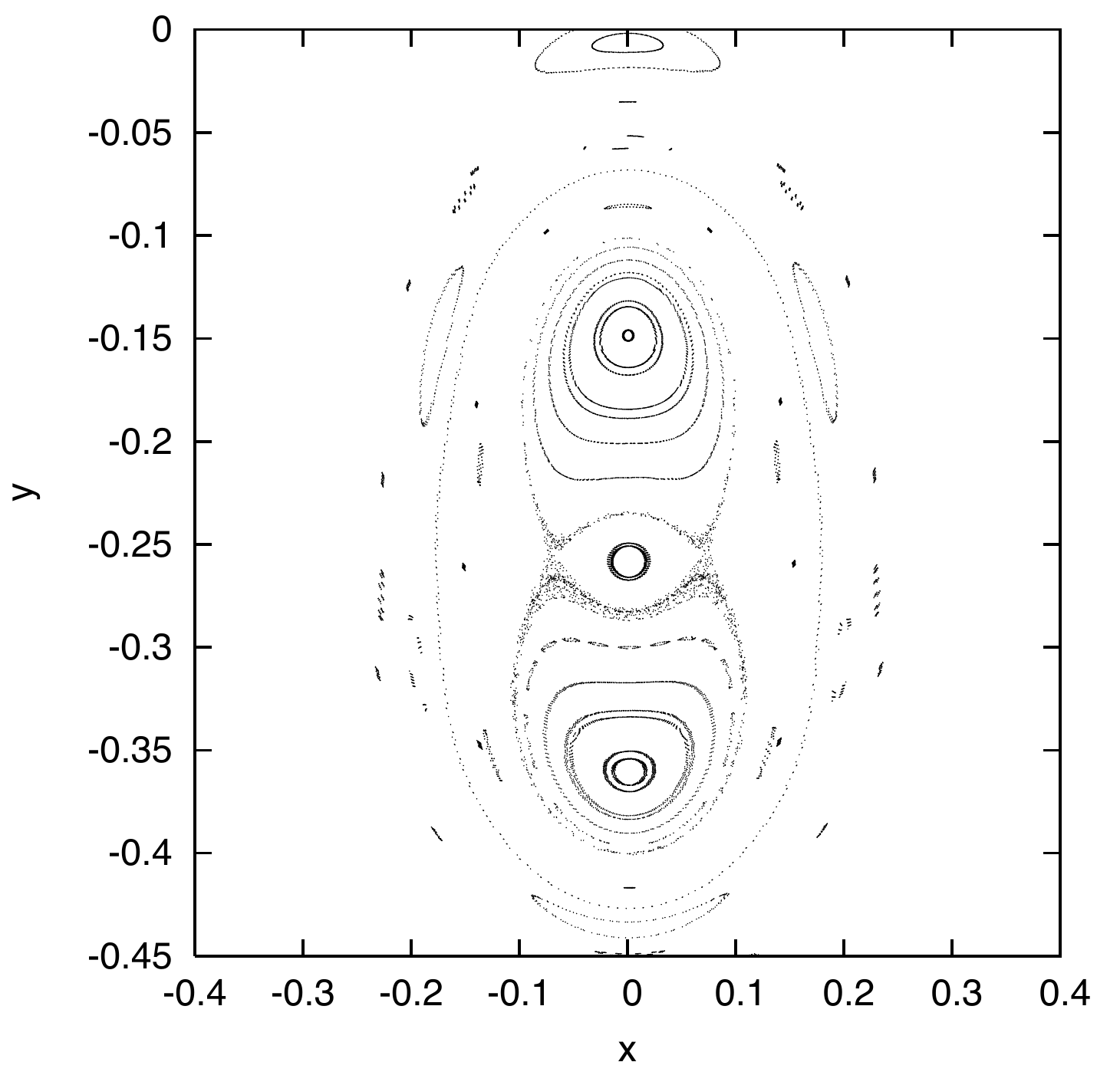}
 {Bifurcation sequence near $p_y = -0.2$ for quartic potential \Eq{Quartic}. (a) $E = 0.21$ single stable SLO, (b) $E = 0.22$ SLO destabilized via subcritical pitchfork bifurcation, (c) $E = 0.23$ SLO restabilized via supercritical pitchfork bifurcation.}{QuarticBifs}{2in}

\subsection{Polar Coordinates}
The polar coordinate construction in the previous example suggests the following 

\begin{lem}\label{lem:polar}
The two degree-of-freedom Hamiltonian
\bel{polarH}
	H = \haf p_r^2 + \frac{\pp^2}{2 r^2} + V(r,\phi) \;,
\ee
where $V$ is smooth, has an SLO $(r(t),\phi_0)$ if
\bel{polarV}
	V = R(r) + P(\phi) S(r,\phi) \;,
\ee
where $P'(\phi_0) = 0$ and $R(r)$ and $S(r,\phi)$ are arbitrary smooth functions.\end{lem}

\begin{proof}
From the equations of motion for \Eq{polarH} in polar coordinates $\ddot \phi = 0$ when
\[
	\frac{\pr V(r,\phi)}{\pr\phi}\vert_{\phi = \phi_0} = 0
\]
with general solution \Eq{polarV}.
\end{proof}

As an example, consider the  \emph{hyper}-\HH family of potentials
\bel{HHH}
	U^{(N)}(r,\phi) =  \frac {r^N}{N} \sin N\phi
\ee
which has $C_N$ symmetry for any positive integer $N$. Including the harmonic potential gives the corresponding Hamiltonian
\[
	H = \haf p_r^2 + \frac{p_{\phi}^2}{2 r^2} + \haf r^2 + \frac{r^N}{N} \sin N\phi
\]
where $\pp = r^2 \dot\phi$. Straightline orbits correspond to zeros of
\[
	\dot p_{\phi} = -\frac{\pr H}{\pr \phi} = -r^N \cos N\phi \;, 
\]
so that $N \phi = n\pi/2$, with $n$ a positive odd integer.
For the \HH system ($N=3$), $\phi_i = \frac16 \pi$, $\frac12 \pi,$ and $\frac56 \pi$, in agreement with \Fig{HHorbits}(a).
For the quartic system ($N=4$), $\phi_i =\frac18 \pi$, $\frac38 \pi$, $\frac58 \pi$, and $\frac78 \pi$ in agreement with \Fig{QuarticOrbits}(a).
The motion on an SLO is given by
\[
	\ddot r + r + (-1)^{(n-1)/2} r^{N-1} = 0 \;.
\]
Note that the orbits with $n = 3 \mod 4$ will be unbounded if their energy exceeds the threshold $E = \frac{N-2}{2N}$, but the orbits with $n = 1 \mod 4$ are bounded for all positive energy values.

Finally, consider the superposition of two hyper-\HH potentials
\be
	U^{(N)} + U^{(M)} = \frac{r^N}{N}\sin N\phi + \frac{r^M}{M}\sin M\phi.
\ee 
By \Lem{Summation}, this potential has SLOs at the common slopes of the two homogeneous potentials, or when 
\[
	\phi = \frac{n\pi}{2N} = \frac{m\pi}{2M} \;,
\] 
for some positive, odd integers $m$ and $n$. Hence, SLO's occur whenever $nM = mN$, an interesting little problem in Diophantine analysis. Since $(n,m)$ are odd it is clear that $N$ and $M$ must both be even or both odd. Thus, a superposition of the \HH potential \Eq{HH} and the quartic \Eq{Quartic} has no SLOs. If $N$ and $M$ are both odd, a simple family of solutions occurs when $n = kM$, $m= kN$ for any natural number $k$. For example, for $(N,M) = (3,5)$, an SLO occurs for $(n,m)=(5,3)$. Note that if $N$ and $M$ have any common factors then these can be removed from the homogeneous equation, and once they are removed the remaining factors of $N$ and $M$ must both be odd (if they were both even, then another factor of 2 can be removed). Thus when $N$ and $M$ are both even then they must have the same power of two in their prime factorization. For example, if $N = 2^2$, then we must have $M = 2^2 (2j+1)$ for some integer $j$. Thus the first common SLOs occur when $M = 12$, for example, with $(n,m) = (k,3k)$.

\subsection{Three Dimensional Examples}
The direct problem \Eq{QN3D} is easily solved for any given potential. For example, the 3D \HH-like model
\[
	V(x,y,z) =  \frac12(x^2+y^2+z^2) + (x^2+z^2)y -\frac13 (y^3 + z^3) \;,
\]
has two SLOs:
\[
	Sl(H) = \left\{ (\tfrac{1}{\sqrt{3}},0), (-\tfrac{1}{\sqrt{3}},0) \right\} \;.
\]
Similarly, the potential
\[
	V(x,y,z) = \frac12(x^2+y^2+z^2) + x^3+y^3+z^3+xyz
\]
has five SLOs:
\[
	Sl(H) = \{(0,0),( \tfrac12,1),(1,\tfrac12),(1,1),(2,2)\} \;.
\]

Since for any degree, \Eq{QN3D} can be viewed as just two equations for the coefficients of $V$ as a function of $\alpha$ and $\beta$, there are many solutions of the inverse problem. For example, the cubic potential
\[
	U^{(3)} = \frac{1-2\beta^2 -2\alpha^2}{3\alpha}x^3 + yx^2+\frac{\beta}{\alpha}x^2z \;.
\]
has an SLO $y=\alpha x$, and $z=\beta x$. When $\alpha = \beta = 1$ this reduces to 
\bel{cubicEx}
	U^{(3)} = -x^3 + (y + z)x^2 \;.
\ee
This potential has an additional SLO, that can be found from \Eq{QN3D}, so that its slope spectrum is
\[
	Sl(H) = \{(1,1),(1,-\tfrac14)\} \;.
\]
A contour plot of $U$ for $E = 0.1$, including the SLO's is shown in \Fig{cubic}. 

\InsertFig{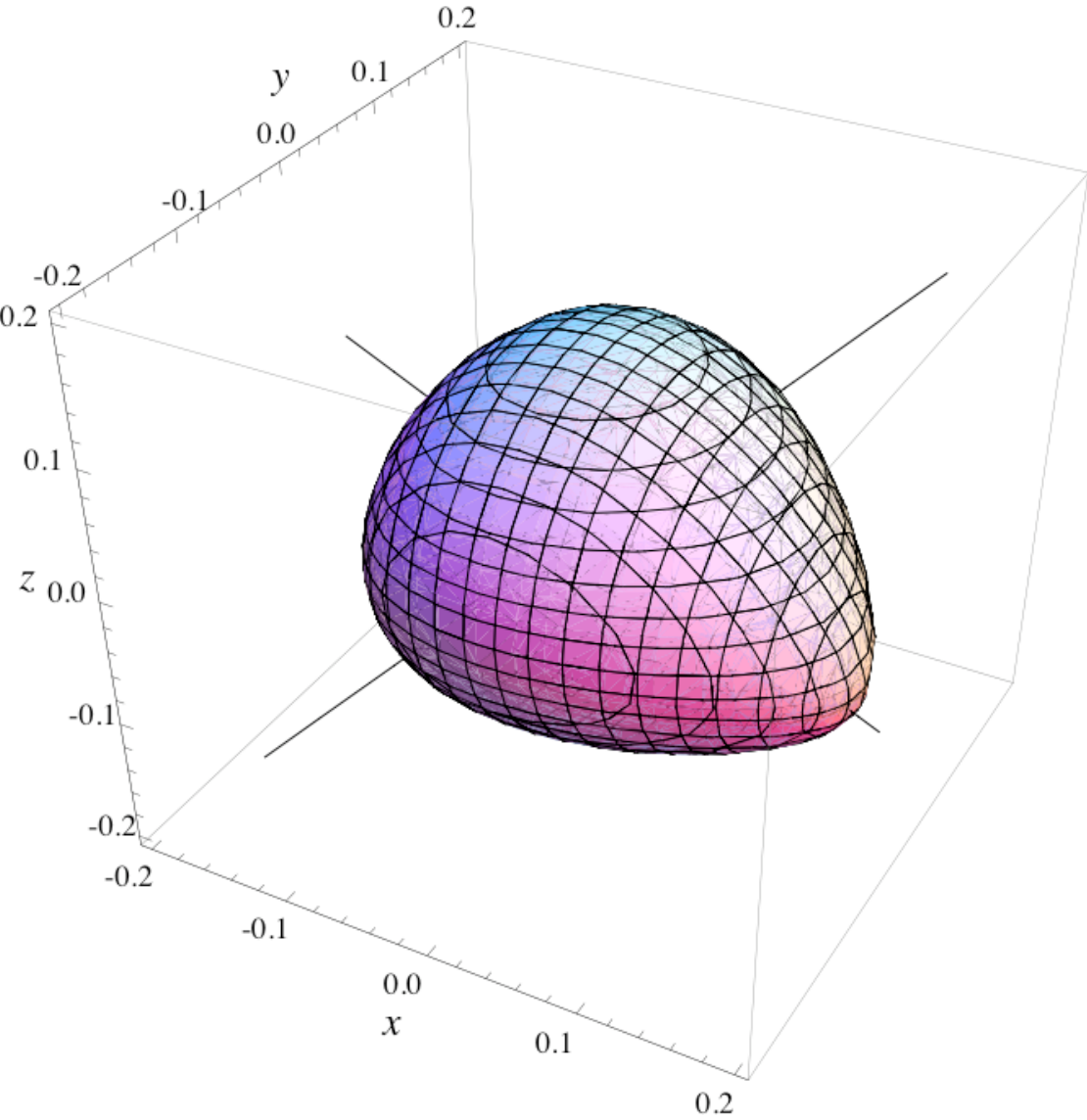}{Contour plot of the cubic potential \Eq{cubicEx} for $E = 0.1$, showing straight line orbits.}{cubic}{3in}

A quartic solution of \Eq{QN3D} is
\[
	U^{(4)} = x^4 + \frac{1}{\alpha^2} y^4 + \frac{1}{\beta^2}z^4 \;.
\]
So, we can get a nice example, we compact energy surfaces, and presumably chaotic orbits by putting these together
\[
	H(x,y,z,p_x,p_y,p_z) = \frac12(p^2 +x^2 +y^2+z^2) + U^{(3)}(x,y,z) + U^{(4)}(x,y,z).
\]
Since the quartic terms dominate for large coordinates, they bound the motion.

As in the 2D problem one can readily incorporate multiple SLOs in the inverse problem. For example, from \Cor{3D}, the potential
\[
	V(x,y,z) = x^2 F_1(x,y,z) +G_1(y,z) 
	         = y^2 F_2(x,y,z) + G_2(x,z) 
	         = z^2F_2(x,y,z)+ G_3(x,y).
\]
has three SLOs along the coordinate axes.
That is, we want the equations of motion for each variable to be of the form
$ \ddot{x} = xK(x,y,z)$, so that $x = 0$ is a solution. An example is
\[
	V(x,y,z) =  \alpha x^2+ \beta y^2+\gamma z^2 + a x^4 + by^4 + cz^4 + 
				dx^2 y^2 + ex^2 z^2 + fy^2z^2.
\]

\section{Discussion}\label{sec:Discussion}

We have determined very general conditions for SLOs for natural potentials in arbitrary dimension. Using these results one can either construct potentials 
with a given SLO or test a given potential for SLOs. The general solution for two degrees of freedom involves two arbitrary functions, for three degrees of
freedom, four.  Superpositions of potentials having SLOs are also easily constructed.  The special case of homogeneous functions occurs rather frequently and as examples we studied the \HH system and a family of hyper-\HH systems, in polar coordinates.  Several two- and three-dimensional examples have been analyzed.

It may be possible to apply similar methodology to more general problems, e.g., to find all potentials with quadratic orbits and to generalize the form of the Hamiltonian to include a mass matrix. It would be interesting to learn whether similar behavior also occurs in non-Hamiltonian, reversible systems.

\paragraph {Acknowledgments}  One of us (JEH) is grateful to George Bozis for many helpful discussions during a very pleasant trip to Thessaloniki. We would also like to thank Professor Bozis for sharing some unpublished work on straight line orbits that was helpful in checking our results.
JEH was supported in part by the Cassini project and JDM by NSF grant DMS-0707659.

\end{document}